\documentclass[aps,prl,twocolumn,showpacs,superscriptaddress]{revtex4}
\usepackage{amsmath,amssymb} 
\usepackage{graphicx}
\usepackage{epsfig}

\renewcommand{\a}{\alpha}
\renewcommand{\b}{\beta}

\newcommand{\s}{\sigma}
\newcommand{\D}{\Delta}

\renewcommand{\dag}{\dagger}

\newcommand{\BK}[1]{\left[#1\right]}
\newcommand{\bk}[1]{\left(#1\right)}

\newcommand{\lr}[1]{\langle#1\rangle}

\newcommand{\be}{\begin{equation}}
\newcommand{\ee}{\end{equation}}
\newcommand{\ba}{\begin{eqnarray}}
\newcommand{\ea}{\end{eqnarray}}

\begin{document}
\title{Proposed realization of itinerant ferromagnetism in optical lattices}
\author{Shizhong Zhang}
\affiliation{Department of Physics, University of Illinois at
Urbana-Champaign, Urbana, Illinois, 61801-3080.}
\author{Hsiang-hsuan Hung}
\affiliation{Department of Physics, University of California, San Diego,
CA 92093}
\author{Congjun Wu}
\affiliation{Department of Physics, University of California, San Diego,
CA 92093}
\date{\today}

\begin{abstract}
We propose to realize the itinerant ferromagnetism of two-component cold 
fermionic atoms in the $p$-orbital bands in optical lattices.
The band flatness in the two-dimensional honeycomb lattice dramatically
amplifies the interaction effect driving the ferromagnetic transition
even with a relatively weak repulsive interaction.
This scheme has the advantage that the stability of the system can be
maintained  without suffering decaying 
to the molecular state as one approaches the 
Feshbach resonance from the side with positive scattering length.
Experimental signatures and detections are also discussed.
\end{abstract}
\maketitle

Itinerant ferromagnetism(FM) is one of the central topics in condensed matter 
physics \cite{Hertz1976, Millis1993, belitz2005, 
lohneysen2007, hirsch1989, moriya1975, mattis1981},
whose mechanism remains a controversial problem. 
The major difficulty is that FM does not have a well-defined
weak coupling picture, i.e., spontaneous spin polarization 
requires interactions strong enough to overcome the kinetic energy cost.
Furthermore, unlike most superconductors in which fermionic 
excitations are gapped except at nodal points which are with zero measure,
ferromagnets have entire gapless Fermi surfaces.
The gapless particle-hole excitations 
dramatically complicates the nature of the transition.
An important progress in understanding ferromagnetism has been made by 
Mielke \cite{mielke1993} and Tasaki \cite{tasaki1995},
who provide a class of exactly solvable models with FM ground states.
The key point is the existence of Wannier-like orbitals for single 
particle eigenstates which reduces the kinetic energy cost of spin 
polarization to zero. However, these  models suffer the 
stringent conditions of fine-tuned long range hopping, which are 
difficult to realize.

In spite of its importance, itinerant ferromagnetism has not been 
realized in cold atom systems partly because of the requirement of 
strong repulsion mentioned above.
It has been proposed \cite{duine2005} to achieve ferromagnetism 
by approaching the Feshbach resonance from the side with positive 
scattering length.
However, a problem is that before the interaction is strong 
enough to drive the transition, the system is unstable toward to the 
formation of the dimer-molecules \cite{ho2004}.

On the other hand, orbital physics in optical lattices  has  received a 
great deal of attention, which
gives rises a variety of new states of matter with both cold 
bosons and fermions \cite{isacsson2005,liu2006,kuklov2006,wu2007,wu2007a,
stojanovic2008,xu2007,wuzhai2007,mueller2007,sebby-strabley2006}.
In particular, it has been recently shown that the $p$-orbital band
in the honeycomb lattice, a $p_x(p_y)$-orbital counterpart of graphene,
has a natural excellent flat band structure.
The consequential non-perturbative interaction effects 
(e.g. Wigner crystallization) associated with the band flatness
have been investigated \cite{wu2007, wu2007a}.
Experimentally, the honeycomb lattice has been constructed quite some 
time ago \cite{grynberg1993}.

In this article,  we propose to realize the ferromagnetic state in
the $p$-orbital band of the honeycomb optical lattice by taking
advantage of its band flatness. 
Due to the divergence of the density of states, even weak repulsions 
can drive the ferromagnetic transition and ensures the system stability 
escaping from the formation of dimer molecules, thus overcome this 
experimental difficulty.
We show the existence of the exact ferromagnetic ground state 
at the filling level $0.25<\lr{n}<0.5$,
and investigate the phase boundaries
between ferromagnetic, paramagnetic, and antiferromagnetic states 
through the self-consistent mean-field method. 
This research will open up a new opportunity to investigate ferromagnetism 
with precise controllability.

We begin with the free part of the 
$p_x (p_y)$-orbital band Hamiltonian in the honeycomb lattice
\ba
\label{eq:ham0}
H_0&=&t_{\|}\sum_{\vec{r}\in A, i, \sigma}\bk{p^\dag_{\vec{r},i,\s}
p_{\vec{r}+a\hat{e}_i,i,\s}+h.c.} -\mu\sum_{\vec{r}\s}n_{\vec{r}\s}
, \ \ \ 
\ea
where $\hat{e}_{1,2}=\pm\frac{\sqrt{3}}{2}\hat{e}_x+\frac{1}{2}\hat{e}_y$
and $\hat{e}_3=-\hat{e}_y$ are the unit vectors pointing from an $A$-site to
its three $B$-site neighbors; $p_i\equiv
(p_x\hat{e}_x+p_y\hat{e}_y)\cdot \hat{e}_i~(i=1\sim 3)$
are the projections of the $p$-orbitals along the $\hat e_i$ direction;
$n_{\vec{r},\s}$ is the number operator for spin $\s$; 
$\mu$ is the chemical potential.
As illustrated in Ref. \cite{wu2007, wu2007a}, its band structure 
contains both flat bands and Dirac cones
whose spectra are symmetric respect to the zero energy.
Two dispersive bands in the middle have two non-equivalent Dirac points
with a band width of $\frac{3}{2} t_\parallel$.
The bottom and top bands turn out to be completely flat in the absence 
of the $\pi$-bonding $t_\perp$, which means that the single particle 
eigenstates of Eq. \ref{eq:ham0} can be constructed as a set of 
degenerate \emph{localized} states.
There exists one such eigenstate in the bottom band for each hexagon 
plaquette, whose orbital configuration on each site is along the 
tangential direction around the plaquette.
The anisotropy of the $\sigma$-bonding and the destructive 
interference together forbid the particle ``leaking'' outside,
which renders these states eigenstates.

The interaction of the spinful neutral fermions is described within the 
$s$-wave scattering approximation, leading to a two-band 
Hubbard model constructed as 
\ba\label{eq:inter}
H_{int}&=&U\sum_{\vec{r}}(n_{\vec{r},x\uparrow}n_{\vec{r},x\downarrow} +
n_{\vec{r},y\uparrow}n_{\vec{r},y\downarrow})-J\sum_{\vec{r}} 
(\vec S_{\vec{r}x}\cdot
\vec S_{\vec{r}y} \nonumber \\
&-&\frac{1}{4}n_{\vec{r}x}n_{\vec{r}y})+\D(p^\dag_{\vec{r}x\uparrow}
p^\dag_{\vec{r}x\downarrow}p_{\vec{r}y\downarrow}p_{\vec{r}y\uparrow}
+h.c.)  
\ea
where $U=\frac{4\pi\hbar^2}{m}a_s\int |p_x({\vec r})|^4 d^3{\vec r}$
($a_s$ the scattering length),
$J=\frac{2U}{3}$, $\D=\frac{U}{3}$, $\vec S_{\vec{r}x}=\frac{\hbar}{2}
p^\dag_{{\vec{r}x\a}}\vec \s_{\a\b}p_{\vec{r}x\b}$ is the spin operator at
site $\vec{r}$ in the $p_x$-orbital, and $\vec S_{\vec{r}y}$ can be
defined accordingly.
The $U$-term is just the usual Hubbard term;
the $J$-term represents Hund's rule physics: the on-site repulsion 
between two fermions in the spin triplet states is zero, while that
of spin singlet is $J$; the $\D$-term describes the pair scattering 
process between the $p_x$ and $p_y$-orbitals on the same site.
We first consider two fermions on the same site
to gain some intuition.
If they are in the spin triplet channel, their orbital wavefunction is
anti-symmetric, thus the $s$-wave interaction vanishes.
If they are in the spin singlet channel, their orbital wavefunctions
are symmetric as $p_x^2+p_y^2$, $p_x^2-p_y^2$, and $p_x p_y$, 
respectively.
The first one has energy $U+\Delta=\frac{4U}{3}$,
while the later two are degenerate with energy $U-\Delta=J=\frac{2U}{3}$.
 
\begin{figure}
\centering\epsfig{file=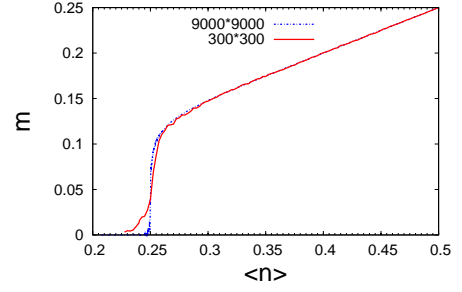,clip=1,width=0.7\linewidth,angle=0}
\caption{(color online)
The magnetization $M$ v.s. $\langle n \rangle $ for $0<\langle n \rangle
<\frac{1}{2}$ simulated on a triangular lattice with $9000\times 9000$
lattice sites (blue line) and $300\times 300$ lattice sites
(red line). Note the strong finite size effect around the percolation
threshold $\lr{n}_c=1/4$.}
\label{fig:percolation}
\end{figure}

At low filling factors, $\lr{n}\le \frac{1}{2}$, the ground states are
heavily degenerate. 
Each fermion occupies one localized plaquette state.
The touching of two neighboring plaquettes with the same spin costs
zero repulsion energy, while that of the opposite spin costs energy
of $\frac{U}{36}$.
According to the percolation picture of the flat band ferromagnetism
\cite{mielke1993,tasaki1995}, fermions form disconnected clusters
composed of touching plaquettes.
Let us denote the total number of sites $N$. There are then $\frac{N}{2}$
plaquettes whose centers form a triangular lattice.
These clusters are labeled from 1 to $n_c$. The number of plaquettes 
inside the $i$-th cluster is denoted as $m_i ~(i=1\sim  n_c)$.
Thus the particle density is just $\langle n\rangle=\sum_{i=1}^{n_c} m_i/N$.
Within each cluster fermions are fully polarized, but the polarizations
of different clusters are uncorrelated.
The ferromagnetic order parameter when $\lr{n}\le \frac{1}{2}$ 
is defined as 
$
M= \sqrt{\langle S^2_{z,tot}}\rangle/N,
$
where the square of total spin of the system is just
$S^2_{z,tot}=(\frac{\hbar}{2})^2\sum_{i=1}^{n_c} m_i^2$,
$\langle \rangle$ means average over all possible
configurations of the degenerate ground states.
As the filling increases beyond the percolation threshold, clusters
become connected driving the system from paramagnetic state to 
ferromagnetic state.
We have performed numeric simulations and obtained $M$ {\it v.s.}
$\langle n \rangle$ as depicted in Fig.\ref{fig:percolation}.
$M$ becomes nonzero approximately beyond the percolation 
threshold $n_c=0.25$, (the small deviation is due to finite size effect),
which corresponds to the site percolation threshold $p_c=0.5$ for
the triangular lattice. 
As we go beyond the percolation threshold, there is an unique infinite
cluster with non-zero probability in triangular lattice 
\cite{strauffer1994}, thus $M$ is just the probability of a site belonging
to the infinite cluster and behaves as $M\propto(\lr{n}-n_c)^\beta$
with the critical exponent $\beta=\frac{5}{32}$ \cite{strauffer1994}.
As $\lr{n}$ increases further, $M$ is linear with $\lr{n}$, 
which means that the system is almost fully polarized.

After the lowest flat band is fulfilled with spin polarized fermions,
{\it i.e.}, $\lr{n}>\frac{1}{2}$, an exact solution is no longer available.
We perform a self-consistent mean-field calculation. 
To facilitate the decomposition of the interaction term of Eq. \ref{eq:inter},
we can recast it in an equivalent form
$H_{int} = \sum_{\vec{r}}\frac{8U}{3}I_{\vec{r}\uparrow} 
I_{\vec{r}\downarrow}
+\sum_{\vec{r}}\frac{4U}{3}(\tau^{(1)}_{\vec{r}\uparrow}
\tau^{(1)}_{\vec{r}\downarrow}+\tau^{(2)}_{\vec{r}\uparrow}
\tau^{(2)}_{\vec{r}\downarrow})$,  
where the pseudo-spin operators $\tau^{(1, 2)}_{\uparrow,\downarrow}$ 
are defined as
$\tau^{(1)}_{\uparrow,\downarrow}=\frac{1}{2}(p_{x\uparrow,\downarrow}^\dag
p_{y\uparrow,\downarrow}+p_{y\uparrow,\downarrow}^\dag 
p_{x\uparrow,\downarrow}),
\tau^{(2)}_{\uparrow,\downarrow}=\frac{1}{2}(p_{x\uparrow,\downarrow}^\dag
p_{x\uparrow,\downarrow}-p_{y\uparrow,\downarrow}^\dag 
p_{y\uparrow,\downarrow}),
I_{\uparrow,\downarrow} = \frac{1}{2}(p_{x\uparrow,\downarrow}^\dag
p_{x\uparrow,\downarrow}+p_{y\uparrow,\downarrow}^\dag 
p_{y\uparrow,\downarrow})$, respectively.
Here $\tau^{1,2}$ measures the preferential occupation of
$x$-and $y$-orbitals, $I$ is half of the identity
operator in the $p_x$ and $p_y$ orbitals.  
Although this does not make the spin $SU(2)$ symmetry explicit, 
it is convenient for the mean-field decoupling if the
spin quantization axis is chosen along the $z$-direction.
The mean-field Hamiltonian then reads, apart from a constant part,
\begin{eqnarray}
H_{int}^{mf}
&=& \sum_{\vec{r}}\frac{8U}{3} \big\{ I_{\vec{r}\uparrow}
 \lr{I_{\vec{r}\downarrow}} +\lr{I_{\vec{r}\uparrow}}  I_{\vec{r}\downarrow} 
\big\}+\sum_{\vec{r}}\frac{4U}{3} \big\{ \lr{\tau^{(1)}_{\vec{r}\uparrow}} 
\tau^{(1)}_{\vec{r}\downarrow}\nonumber\\
&+&\tau^{(1)}_{\vec{r}\uparrow}
\lr{\tau^{(1)}_{\vec{r}\downarrow}}+\lr{\tau^{(2)}_{\vec{r}\uparrow}}
\tau^{(2)}_{\vec{r}\downarrow}+\tau^{(2)}_{\vec{r}\uparrow}
\lr{\tau^{(2)}_{\vec{r}\downarrow}} \big\}.
\label{mfham}
\end{eqnarray}
The ferromagnetic order parameter is defined as $M=\sum_{\vec{r}}
\bk{I_{\uparrow}-I_{\downarrow}}/N$. 
We enlarge the unit cell to include six lattice sites to allow
site-dependent magnetic and orbital configurations.
We also replace the bare value of $U$ with an effective one in the
self-consistent calculation by taking into account the Kanamori 
corrections defined as
$U_{eff} (\mu)=  U({\vec p},{\vec q};\mu)$ at $p, q \rightarrow 0$
\cite{Callaway1974, Fazekas1999}. 
$U({\vec p},{\vec q};\mu)$ is defined as
$U({\vec p},{\vec q};\mu)= U/[1-U\Pi{(\vec p},{\vec q})]$,
and 
\ba
\Pi=1/N\sum'_{\vec k}\BK{E({\vec q})+E({\vec p})-E({\vec k})-E({\vec p}
+{\vec q}-{\vec k})},
\ea
where $E(\vec k)$ is the band energy corresponding to Eq.\ref{eq:ham0},
and the summation over $\vec k$ is taken over unoccupied Bloch states. 
The renormalization procedure ensures that $U_{eff}$ can be of the same
order as the band width at most and thus precludes 
ferromagnetism in the usual Hubbard model at low density 
even for very large bare $U$ \cite{Fazekas1999}.

\begin{figure}
\centering\epsfig{file=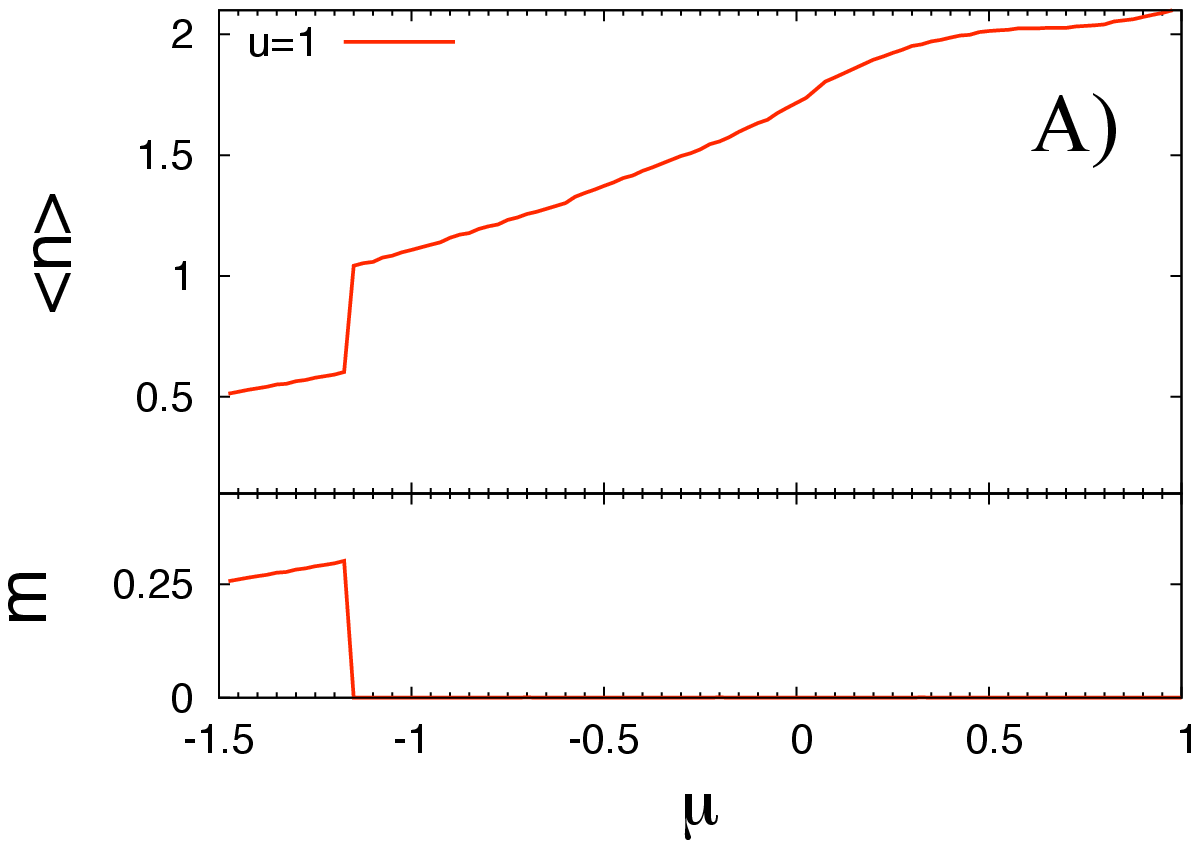,clip=1,width=0.7\linewidth,angle=0}
\centering\epsfig{file=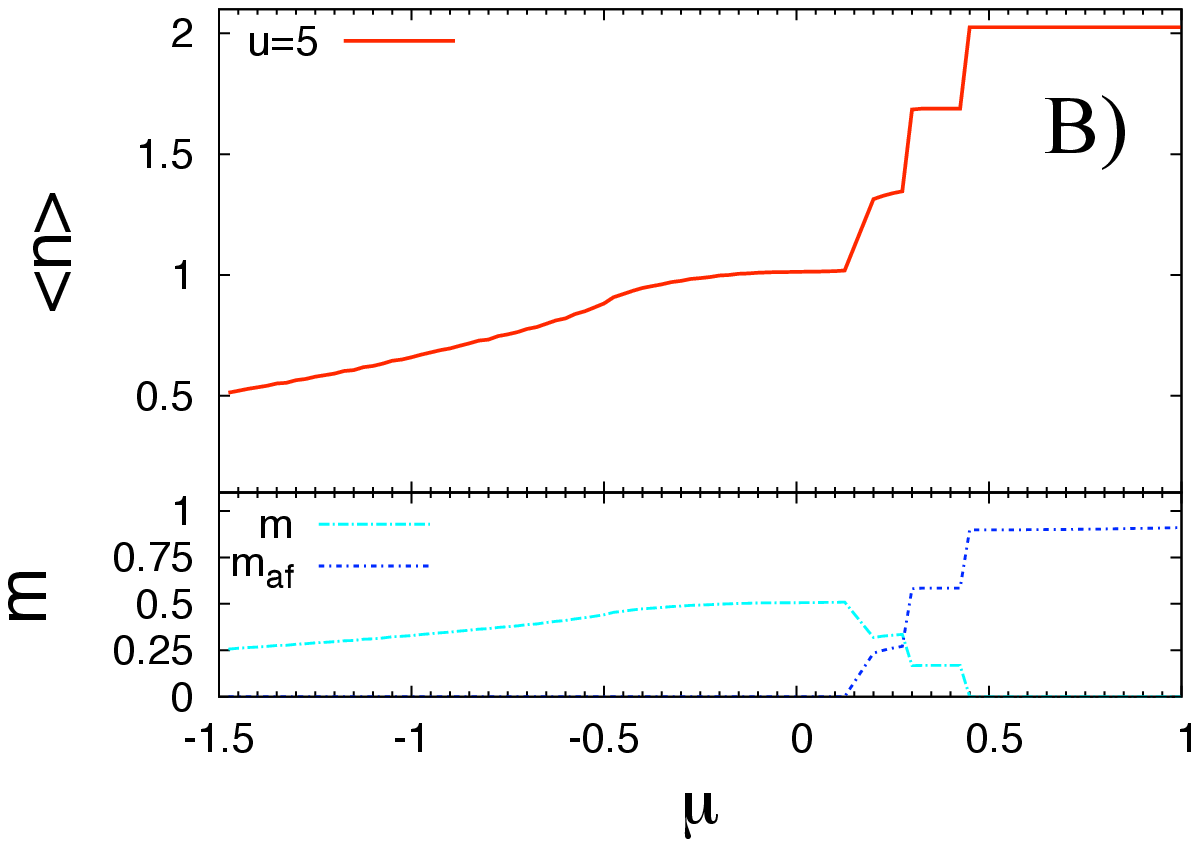,clip=1,width=0.7\linewidth,angle=0}
\caption{(color online) A) Filling factor $n$ and magnetization $m$ v.s
the chemical potential $\mu$ at weak interaction ($U/t_\parallel=1$),
B) $n$, $m$ and antiferromagnetic moment $m_{af}$ v.s. 
$\mu$ at strong interaction ($U/t_\parallel=5$).}
\label{fig:mag}
\end{figure}

As $\lr{n}$ increases beyond $\frac{1}{2}$, the flat band is fulfilled with 
polarized particles with $\mu/t_\parallel>-1.5$.
The additional particles have to either flip their spin and fill the flat
band states costing energy of order $U$, or they continue to fill the 
second band while remaining spin polarized.  
The competition between the kinetic energy and the interaction energy 
determines a critical chemical potential $\mu_c$ as a function of
$U$, beyond which the full polarization is no longer favorable.
Within the mean field theory, we compare the energies of fully polarized 
ferromagnetic states $E_{ferro}(U,\mu)$ and that of a paramagnetic state
$E_{para}(U,\mu)$, and then determine $\mu_c$. 
As depicted in Fig.\ref{fig:mag} A, we show the dependence of 
$\lr{n}$ and $M$ as a function of $\mu$ at a weak coupling 
$U/t_\parallel=1$. 
Note that $\lr{n}$ and $m$ both jump at $\mu_c\approx -1.15$
which is a consequence of the flat band structure.
When the minority spins begin to populate,  the divergent density 
of states of the flat band suddenly increases its filling, 
which significantly changes the effective potential that the majority 
spins feel, and drive the system to paramagnetic states.
In other words, if the particle density is fixed between the two values
before and after the jump, we obtain the phase separation state
as coexistence of the fully polarized state and the paramagnetic state.
After the paramagnetic phase is reached, the system remains in
metallic state at small $U$ because of
the vanishing of density of states of the Dirac cone at half filling.

The phase diagram becomes richer as $U$ goes stronger as depicted
in Fig. \ref{fig:mag} B with $U/t_\parallel=5$.
The fully polarized state survives to a larger $\mu_c/t_\parallel 
\approx 0.1$ as expected; the Mott-insulating antiferromagnetic state
occurs at half-filling $\lr {n}=2$ with the N\`{e}el order defined as 
$ M_{af}=\frac{1}{N} \sum_i (-)^i S^i_z$.
In the strong coupling limit at half-filling, each site is filled with 
two particles with spin aligned by Hund's rule as a spin-1 complex. 
The virtual hopping of particles leads to the antiferromagnetic exchange
and the long range N\`{e}el order in the ground state.
At intermediate fillings between the fully polarized state and 
antiferromagnetic state, as $\mu>\mu_c$, the quick increase of
$\lr{n}$ is due to the filling of the minority spins in the flat band.
Interestingly, an incompressible state occurs at $\lr{n}=5/3$ with
$\lr{n_\uparrow}=1$ and $\lr{n_\downarrow}=2/3$, which has the 
coexistence of the partial ferromagnetic and antiferromagnetic orders,
i.e., the ferrimagnetic order.
It also exhibits intricate orbital ordering pattern as in the
spinless case \cite{wu2007,wu2007a}.
We will defer a detailed analysis of this state to a future publication.

Now we discuss the effect of the small $\pi$-hopping term $t_{\perp}$. 
Then the lowest energy band acquires a weak dispersion with band width 
of order $t_{\perp}$.
In this case, the localized plaquette states are no longer eigenstates 
of the system. 
However, so long as the interaction strength $U$ is much
larger than $t_{\perp}$, it is not energetically favorable to flip the
spin and thus the ferromagnetic state is stable. 
As shown in Ref. \cite{wu2007a}, in realistic systems with 
the sinusoidal optical potential, it is easy to suppress 
$t_\perp/t_\parallel$ to $10^{-2}$ or even $10^{-3}$.
We have confirmed the existence of ferromagnetic order numerically
by using the above mean-field theory for $t_{\perp}$ up to
$0.1t_{\parallel}$. 

Another small interaction which was previously left out 
is the $p$-wave scattering $U'$. It adds an extra interaction
term $\Delta H_{int}=\sum_{\vec{r}} U' n_{\vec{r}x}n_{\vec{r}y}$,
and the value of $J$ in Eq. \ref{eq:inter} changes to
$J'=\frac{2}{3}U-U'$.
In this case, the flat band structure is not destroyed, but an exact
solution is no long available as long as $\lr{n}>1/6$
because different plaquettes touch and cost energy of order
of $U^\prime$ even with polarized spin.
We have also performed self-consistent mean field theory to 
check that the ferromagnetic state is still stable
at small value of $U^\prime$.
For example, for the case of $U/t_\parallel=5,U'/t_\parallel=1$,
we find that the ferromagnetic order survives up to filling 
factor $\lr{n}\approx 1$.

\begin{figure}
\centering\epsfig{file=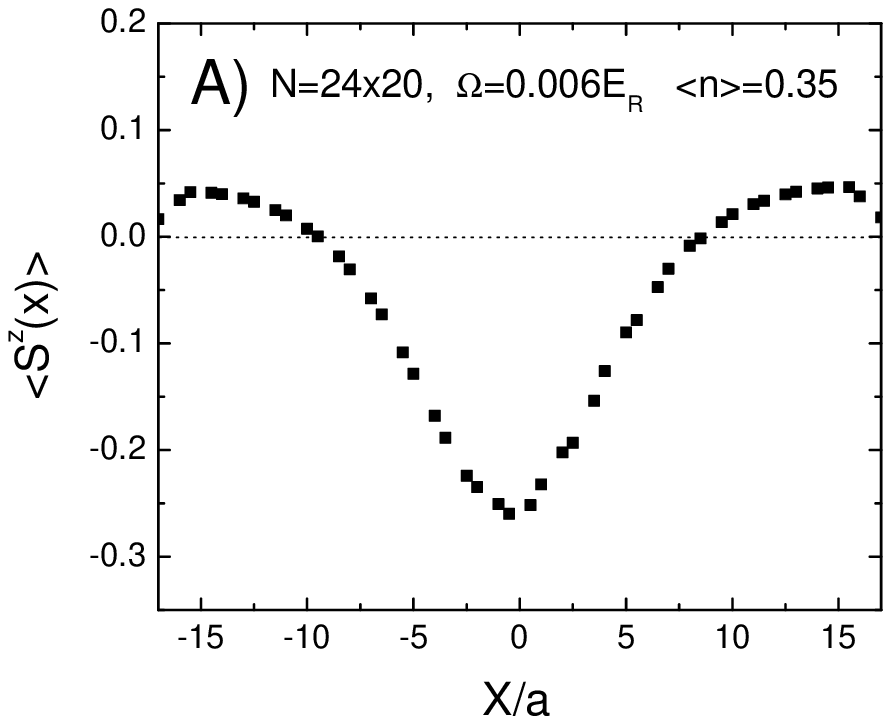,clip=1,width=0.48\linewidth,angle=0}
\centering\epsfig{file=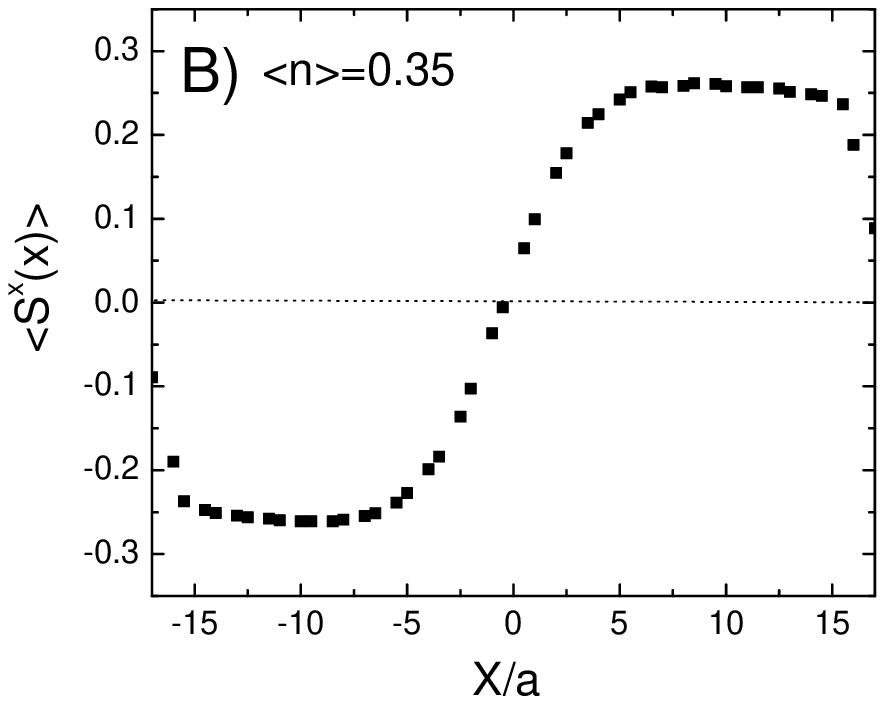,clip=1,width=0.49\linewidth,angle=0}
\caption{
The skyrmion configuration of the spin density distribution 
in the confining trap with parameters $U/E_R=1$, $t_\perp/E_R=0.24$, 
$\Omega/E_R=0.006$, and $n_\uparrow=n_\downarrow=168$.
The total number of unit cells is $N=24\times 20$ 
(the number of sites is $2N$). 
A) $\langle S_z (\vec r)\rangle $ along the line of $(X/a, Y/a=0)$, 
{\it i.e.}, the center row in the system.
B) $\langle S_x(\vec r) \rangle$ pattern along the line of $(X/a, Y/a=0)$. 
$\langle S_y(\vec r) \rangle$ shows the similar behavior along
$(X/a= 0, Y/a)$.}
\label{fig:trap}
\end{figure}

Now we discuss the finite temperature effect.
We only consider the case of the filled flat band whose low energy 
excitations are captured by an effective ferromagnetic Heisenberg 
model defined for plaquette spins \cite{mielke1993}, and leave the 
study of the general filling to a future publication.
In our case, the effective model is constructed on a triangular lattice as
$H_{ex}= -J^\prime \sum \vec S_i \cdot \vec S_j$ with 
$J^\prime \approx \frac{U}{18}$.
The mean field calculation based on Eq. \ref{eq:inter} shows
that the critical temperature $T_{mf}$ is at the same order of 
$zJ^\prime/2$ for small values of $U/t_\parallel$ at 
$\langle n \rangle =\frac{1}{2}$ where $z=6$ is the coordination number.
$T_{mf}$ is the temperature scale for the onset of the nonzero 
magnitude of the ferromagnetic order parameter, which is roughly one 
order smaller than $U$.
According to the data in Ref. \cite{ho2007}, $T_{mf}$ can be estimated 
as the order of $1\sim 10$ nK which remains experimentally accessible.
Below $T_{mf}$, thermal fluctuations of the orientation of the ferromagnetic 
order are described by the $O(3)$ non-linear-$\sigma$ model, and no
true long range order exists for $T>0$ in 2D.
However, the spin correlation length $\xi$ diverges fast as 
$\xi/a\approx e^{T_{mf}/T} $ ($a$ is the lattice constant) 
which can easily exceed the small size of the optical lattice 
(e.g. typically $L/a \approx 50$) as $T$ goes smaller than $T_{mf}$.
Thus practically we can still take $T_{mf}$ as the onset temperature
scale of ferromagnetism.

Next we study the effect of the overall harmonic trapping potential
$V_{tr}(\vec r)= \frac{m}{2} \Omega^2 r^2 $ and show that ferromagnetism 
remains robust through a self-consistent Bogoliubov-de Gennes 
calculation in real space.
In experiments, the particle number of each component 
is separately conserved.
Furthermore, when the system has $SU(2)$ symmetry as in Eq. \ref{eq:inter},
the spin configuration prefers the skyrmion configuration rather 
than the Ising domain wall \cite{berdnikov2009} to lower the energy.
The skyrmion-type real space magnetization distributions of
$\langle S_z(\vec r) \rangle$  and $\langle S_x(\vec r)\rangle$
are depicted in Fig. \ref{fig:trap} A and B, respectively,
with the parameters $U/t_\parallel=1$; $\Omega= 0.006 E_R$ 
($E_R$ the recoil energy);  $t_\parallel/E_R
=0.24$ as calculated in Ref. \cite{wu2007a}.
The characteristic trap length scale is $l=\sqrt{\frac{\hbar}
{m\Omega}}\approx 4.1a$ where $a$ is the lattice constant.
We take particle numbers of spin $\uparrow$
and $\downarrow$ components as
$n_\uparrow=n_\downarrow=168$ which can be achieved 
by introducing two different chemical potentials for
$\mu_\uparrow=-1.3897$ and $\mu_\downarrow=-1.3903$.
The spin $\downarrow$ particles concentrate in the center region,
whereas most spin $\uparrow$ particles live outside and form a ring. 
The spin distribution smoothly varies from $\downarrow$ to 
$\uparrow$ from the central region to the outside 
due to the spin $SU(2)$ symmetry.
Experimentally, the shell structure can be resolved {\it in situ} by 
tomographic radio-frequency spectroscopic method \cite{Shin2007} 
followed by a phase-contrast imaging technique \cite{Shin2006}
which can distinguish the shell structures with
two different components.

In summary, we have proposed a scheme to realize the itinerant 
ferromagnetism by using the flat band structure in the $p$-orbital
system in the honeycomb lattice.
The band flatness stabilizes ferromagnetism even with weak repulsive 
interactions, thus the stability problem of using Feshbach type
methods is avoided.
The ferromagnetism is robust against soft trapping potentials,
exhibiting the skyrmion type configuration as a result of
spin conservation.
Furthermore, due to the large configuration space of the partially
filled flat band, flat band ferromagnetism supports
large entropy which makes this proposal even more realistic. 
This work potentially provides a way to realize spin transport and
even spintronics in the future research with cold atoms.

S. Z. is supported by NSF-DMR-03-50842.
C. W. thanks J. Hirsch and L. Sham for helpful discussions.
C. W. and H. H. Hung are supported by  
the Sloan Research Foundation, ARO-W911NF0810291 and NSF-DMR-0804775.

{\it Note added}\ \ \
After this work was completed, we learned independent works
by Wang {\it et al.} \cite{wang2008} and Berdnikov {\it et al.}
\cite{berdnikov2009} on the ferromagnetism with cold atoms.
Evidence of ferromagnetism has been observed in a recent experiment 
through Feshbach resonances by Jo {\it et al.}\cite{jo2009}


\end{document}